**Who are Political Retweeters?, Demographic comparison of political retweeters with retweeters of non-political personalities.**

Gurchani, Muhammad Umer

Université de Montpellier

muhammad-umer.gurchani@etu.umontpellier.fr

**Conflict of Interest:** The author declares that they have no conflict of interest

**Abstract**

Twitter has been a focus of research in political science for a few years now as it provides the opportunity to make direct observations on spread of political information in different communities. Here we will be studying the phenomena of information diffusion, and focus on nodes that are responsible for spreading political information everywhere on the Twitter network. This paper attempts to fill gaps in literature regarding the demographics of political retweeters using various techniques on name and location related data from most active French political retweeters. Here I will try to state the break-down of these accounts in categories based on gender, language, location, education-level, and self-descriptions. To put the information about political retweeters in context we will also create a category of non-political retweeters to draw comparisons between the groups regarding the above-mentioned variables.

**Introduction**

In the recent decade, retweeting has become one of the most important information diffusion mechanism for politicians. Many major political figures in the world tweet frequently to stay in touch with their followers and try to engage with them on a regular basis. Some of these tweets get retweeted more often than others. In order to understand why people, retweet and how the retweeted information defuses in a network it is important that we first investigate into who retweeters are, and group them into meaningful demographic categories that can be studied to draw conclusions about spread of political information on Twitter. The main objective of this paper is to put retweeting behavior in context by knowing more about the people behind the accounts who retweet major French politicians. From this inquiry, it has been found that although high-frequency political retweeters have significant number of party-dedicated political participants but overall demography of this group is very similar to randomly selected twitter followers of politicians but very different from accounts that are active retweeters of non-political content, which is helpful in establishing that the act of retweeting itself cannot be associated with twitter accounts of particular demographics. Establishing this will provide ample justification for looking into community dynamics of a group on twitter, and the content of original tweets to decipher the motivations behind retweeting.

**Review of Literature**

**Why is this question important?**

Researchers on political polarization patterns have found that political debate on twitter is largely dominated by a small group of people who are highly interested in politics (Bekafigo & and McBride, 2013) (Tumasjan, et al., 2010). It has also been proposed that the people that are politically active on internet in general tend to be the ones that are also actively participating in political activity in real-life (Bekafigo & and McBride, 2013). From Political science and communication studies perspective some important questions can be asked about the nature of groups in which political information spreads quickly and where political argumentation and debates happen. Are there any demographics differences in these groups? Is there any hierarchy in these groups through which political information trickles down? Are these groups representative of any real-world groups? While these questions have not been directly addressed in literature to the best my knowledge but researchers have tried to check if twitter population in general is representative of real-world population. The answer to this question has been found to be largely negative (Mellon & Prosser, 2017). It has been found that twitter users are largely male and highly concentrated in rich urban areas with younger population (Barbera & Rivero, 2014) (Mislove, et al., 2011). While this is an important result and has serious implications for scholars trying to make real-world predictions based on twitter data (Tumasjan, et al., 2010). However, this result also raises questions about subgroups in twitter and if these sub-groups are reflective of the same trend and what kind of effect would the basis of the group have on the demographics of profiles present in it. For example, one can ask how different are the demographics of group of political retweeters compared to groups that retweet non-political personalities like sportsmen or actors. It is hypothesized here that general demographic trend of twitter will be more magnified in community of political retweeters as compared to other non-political groups. In this paper, I will try to investigate into demo graphics of political retweeters and compare the results with other most popular groups of non-political retweeters.

**Methodology related Literature**

Demographics of users on Twitter have been an important topic in the literature surrounding Twitter. As Twitter does not provide clear demographic information about its users except the name, location, and language, it becomes harder to look at the demographic divisions of the users. In literature, there have been a few attempts to determine more knowledge about the users from the information that Twitter provides. I will list these strategies here and then use them to determine the demographics of the target population.

The ways in which, gender and location of a twitter user can be determined have been discussed by various scholars. As per Sloan (Sloan, et al., 2013), the most reliable way to get the gender of Twitter users is to make use of first names and as far as location is concerned the location provided voluntarily by the users, can be considered a reliable measure when used in conjunction with yahoo place finder. For the sake of this paper, we will not be using yahoo place finder, but Google's geocoder service will be used to determine the region, departement and commune, which has a higher accuracy rate due to the ubiquity of android systems and higher usage of Google maps around the world.

Another methodology that will be used to determine the demographic features of retweeters in France is the usage of departement-level (County Level) official data from insee (Institut national de la statistique et des etudes économiques) for inferring information about twitter profiles. Mohammady originally proposed this method and yielded valid results and we will use it to aggregate the education level of political retweeters (Mohammady & Culotta, 2014). However, since we will be detecting associations on population level for this part of the study. Therefore, it would be not accurate to assume that out conclusions will be applicable to individual retweeters. Otherwise, we will be at the risk of committing ecological fallacy.

**Compared to what?**

In order to explain the findings, there is a need to put the found knowledge in context. There are two basic features in accounts that we want to study

1) They are very interested in politics.

2) They retweet politicians frequently.

Following (Yellow boxes) are the possible groups one can study in order to draw meaningful conclusions about the variables, 'political interest' and 'retweet frequency'.

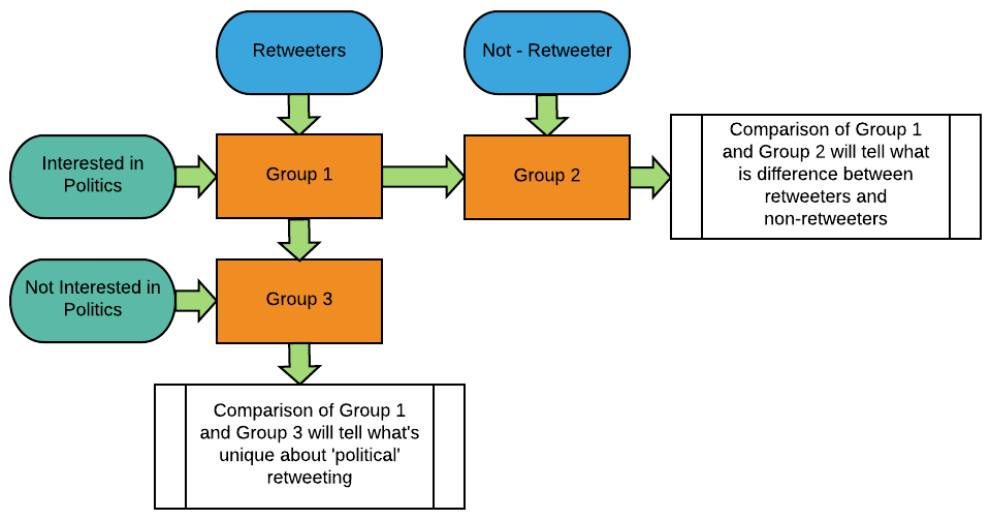

Diagram 1.1: The above diagram shows the 3 groups we will study to draw conclusions about political retweeters.

**Group 1**

The first group that we need to study is the people who are active retweeters of politicians. From the French political context, I selected the top 5 contestants of the French presidential election of 2017. I then found out their official Twitter accounts and the collected their most frequent retweeters using twitter API. The first step towards the process of selecting group 1 will be to sort the retweeters of each of the 5 French politicians based on the frequency of their retweeting. Here the frequency means how often they retweet a politician. The accounts that retweet a politician more frequently were placed on top, followed by others who retweeted less frequently for the same politician. Among these retweeters, I selected only the top 10000 most frequent retweeters of each of the politicians and created 5 tables containing profiles of each of the top 10000 retweeters of politicians. These 5 tables were then merged to create a combined table of 44779 rows containing at least 10000 retweeters of each of the politicians. The reason that there are less than 50000 profiles in this merged database is that many of the top retweeters of one politician were also top retweeters of another politician; therefore, the repeated rows were eliminated in the data cleaning process. In this database of 44779 unique twitter profiles, following is the count of profiles of retweeters who have been retweeting these politicians.

| Politician | Number of Retweeters in the Database |
|---|---|
| Francois Fillon | 14408 |
| Jean Luc Melenchon | 13048 |
| Marine Le Pen | 11914 |

| Emmanuel Macron | 13826 |
|---|---|
| Benoit Hamon | 13914 |

Table 1.1: Politicians we will study

As it can be observed from the table there are considerably more than 10000 values for each of the politicians which shows that retweeting is not an exclusive behavior and an account who frequently retweets one politician can also retweet many other politicians at the same time.

**Group 2:**

The second group will constitute of people who are not active retweeters of politicians, but they follow these politicians, which are assumed as an indicator of their interest in politics. For this group, we selected random 50000 profiles from the list of followers of five major French politicians, repeated the same process for them as for group 1, and collected their profile information. There is a need for this group because using this group, we can observe if some property is specific to retweeters of politicians or is it a general trend among the accounts that are interested in following politicians in twitter.

**Group 3**

The third group will constitute of profiles that are active retweeters but not of politicians. Since the non-political group can be a very large group with multiple subgroups. It was concluded that it would be better to study two non-political subgroups so that found variables can be compared between these sub-groups and then also compared with political retweeters. If there is uniformity in non-political groups and difference with the political group, then it can be concluded that the concerned property is specific to political retweeters group. For this purpose, I selected the most popular personalities in France on Twitter with no explicit political affiliation. Two options were considered, the first one was to use Twitter accounts of soccer players as they are among the most popular Twitter handles in France. The second option was to consider entertainment personalities. It was decided to select both of these groups and study them simultaneously to make meaningful conclusions about non-political retweeters.

| **Sub-Group 1** | **Sub-Group 2** |
|---|---|
| Jamel Debbouze (@debbouzejamel) | Antoine Griezmann (@antogriezmann) |
| Gad Elmaleh (@gadelmaleh) | karim benzema (@benzema) |
| Cyprien (@monsieurdream) | matuidi blaise (@matuidiblaise) |
| Kev Adams (@kevadamsss) | Paul Pogba (@paulpogba) |
| Norman Thavaud (@normandesvideos) | raphael varane (@raphaelvarane) |

Table 1.2 : Non-Political twitter accounts whose retweeters, I will study

A similar database of retweeters was created from these accounts like that of politicians and similarly, detailed information was also collected from these profiles.

**Found Attributes**

The following information was then gathered about the retweeters from the Twitter API, multiple attributes were collected about these retweeters from Twitter API. These attributed include,

**id, name, screen_name, location, description, protected, verified, followers_count, friends_count, listed_count, favourites_count, statuses_count, created_at, time_zone, geo_enabled, lang**

Knowing this information allows us to move toward the analysis of this data set to figure out information that can be useful in studying the general features of political retweeters.

**Language breakdown**

We will start this analysis with the 'lang' attribute which represents the primary language in which a user uses his/her twitter interface. The language attribute is important as it provides an idea of national background of the users and helps us determine if there is linguistic uniformity or diversity among the groups. Here is the percentage breakdown of all the groups in terms of language.

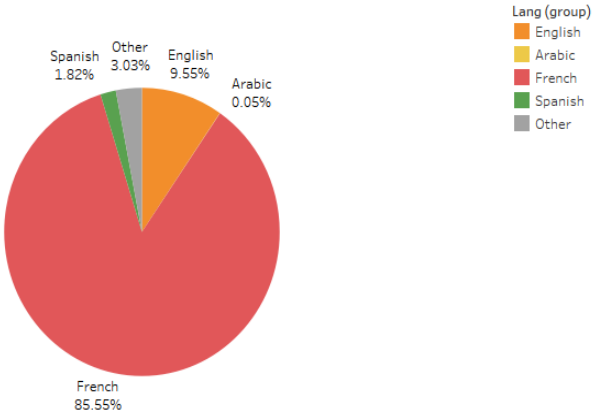

Diagram 1.2 Language breakdown of Group 1 (Political Retweeters)

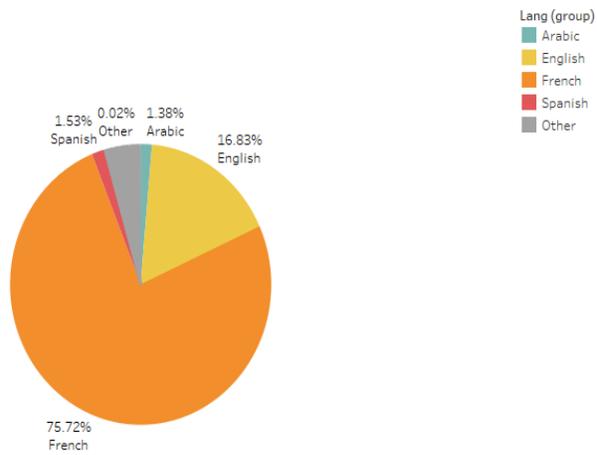

Diagram 1.3 Language breakdown of Group 2 (Political non-retweeters)

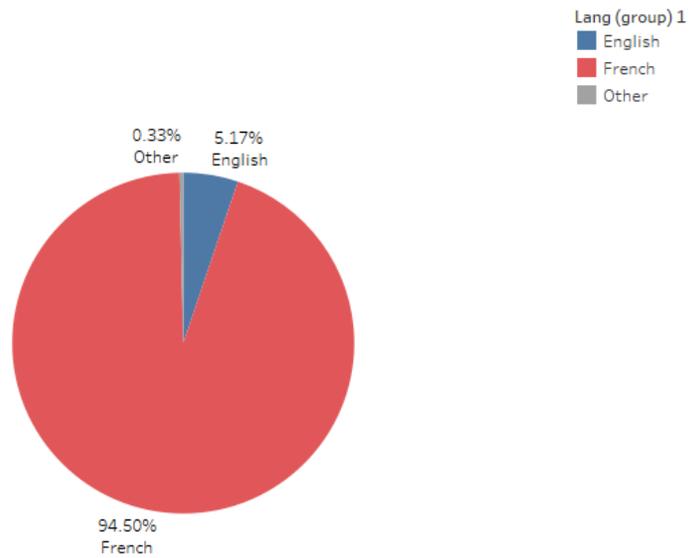

Diagram 1.4 Language Breakdown of Group 3 (Entertainers)

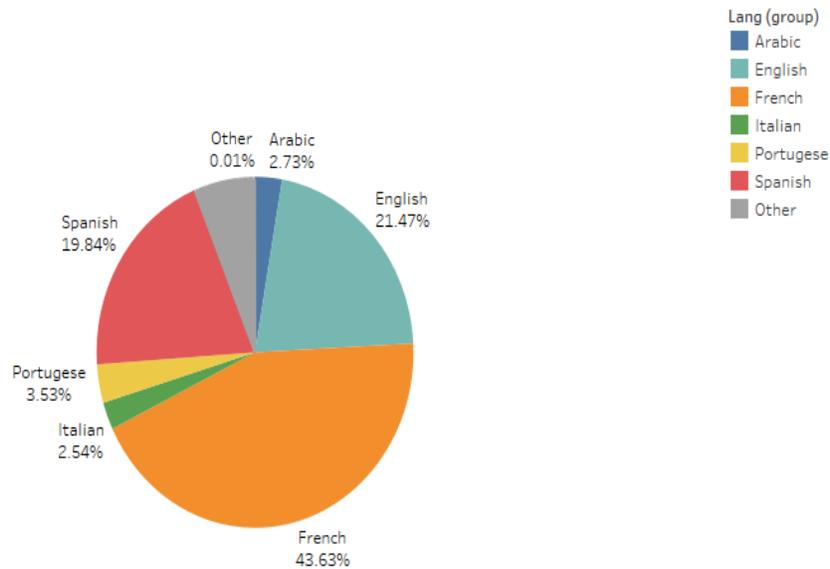

Diagram 1.5: The image above represents the language breakdown of retweeters of footballers

**Discussion about Language break-down of multiple groups**

As seen from the comparison of graphs about the language, the diversity among the language groups is consistent among group 1 and group 2 whereas it is very different from group 3 which has huge variation among the sub-groups as well. This shows that the distribution of languages that we see in group 1 and group 2 is particular to both political retweeters and political non-retweeters and not present in non-political retweeters. This compels to infer that this division is specific to people who are interested in politics.

If we inspect deeper into the division of political retweeter we will see that almost 86 percent of retweeters use the French Language as their interface language in using twitter. While the second most used language is English, which can be explained by the fact that many French people are bilingual. What is noteworthy here is the number of people who have chosen to use Twitter in Arabic. According to data collected from Adult Education Survey in 2007 by European Union, Arabic is the second most common maternal language in France with 3.6 percent of French population speaking it as their mother tongue (European Union Adult Education Survey, 2016). Whereas in the representation of political retweeters, it represents as little as 0.05 percent of the population. There may be several explanations behind this. One possible reason is that although Arabic is a mother tongue for 3.6 percent of people, yet they use French on daily basis and have become fluent enough in French to use websites like Twitter in French rather than their native tongue (Myers, et al., 2014). Another explanation for the lack of Arabic speakers among the retweeters will be the underrepresentation of this ethnic group among the retweeting population of France. Clearly, there is a need to investigate this further and we will do in the next section of this paper where we will break down the first and second names of retweeters to create a classifier to see if there is indeed an under-representation of people with Arabic origin in among the retweeters.

Following is the breakdown of languages for each of the politicians separately:

| LANG (GROUP) | LE PEN | MACRON | MELENCHON | HAMMON | FILLON |
| --- | --- | --- | --- | --- | --- |
| FRENCH | 78.76% | 81.14% | 91.33% | 92.24% | 91.22% |
| ENGLISH | 11.97% | 13.56% | 7.02% | 7.05% | 7.76% |
| ITALIAN | 4.54% | 0.68% | 0.33% | 0.20% | 0.20% |
| OTHER | 1.69% | 1.24% | 0.85% | 0.27% | 0.30% |
| RUSSIAN | 1.68% | 0.27% | 0.05% | 0.00% | 0.08% |
| GERMAN | 0.88% | 0.58% | 0.33% | 0.16% | 0.15% |
| JAPANESE | 0.31% | 0.25% | 0.06% | 0.01% | 0.04% |
| SPANISH | 0.12% | 2.13% | 0.04% | 0.07% | 0.19% |
| ARABIC | 0.04% | 0.15% | 0.00% | 0.00% | 0.05% |
| TOTAL | 100 % | 100 % | 100 % | 100 % | 100 % |

Table 1.3: Percentage of each language for each politician among his/her retweeters

If we look at the table closely, we can see that there are many observations, which require an explanation. The first observation is a follow-up on the lack of Arabic speaking retweeters. Here we can note that the small number of Arabic retweeters that do retweet French politicians prefer to avoid leftist candidates such as Melenchon (0.0 %) and Hammon (0.0 %). This phenomenon needs to be explored further to see the reasons why it is happening.

The second observation is the international appeal of both macron and marine le-pen in contrast to their popularity among French-speaking retweeters. This can be explained by the fact that they reached the second round of French presidential elections of 2017 which had more international media coverage than the first round which included the other 3 candidates too.

The third observation is that non-French speaker appeal of le-pen and macron comes from completely different language groups. While Macron has managed to get retweeted significantly more by Spanish speaking profiles, Marine Le-pen has attracted significantly more Italian and

Russian speakers which can be explained by her general popularity in those countries. (Batchelor, 2017)

**Section 2 (Location Data):**

In this section, we will investigate the location data of retweeters and try to see what kind of areas they come from. We will start with the country level analysis and then concentrate on France, its regional data and data of its departments. We will then use multiple educational and social indicators from the official on ground surveys done by French government organizations to try to see what kind of areas retweeters come from.

**Where does the location data come from?**

The location data of retweeters is taken from the location tab in their profile. This location is self-reported and serves as the best option to approximate the location of these users. The cleaning of this data is much more complicated as location data is not reported in a uniform format. Some profiles just add country names, whereas there are other profiles who have reported their full location. There is also a problem of obviously misreported locations as there are many users who have reported their location to be something like 'dans la lune'[1], 'Narnia' or 'dans la l'espace'[2]. These profiles had to be discarded during this analysis as it was not possible to determine their location. Another problem in the database is that although the majority (approximately 60 percent) of the retweeters have mentioned their location yet there is a significant portion of retweeters who have not revealed their location and it is difficult to ascertain where they are from.

The reported locations were then further broken down in the following columns:

| |
|---|
| Pays/Country |
| Region/State |
| Département/County |
| Arrondissement /Locality |
| Code Postal/ Postal Code |

As twitter only provides a single column for location and this location needed to be cleaned and distributed in above-mentioned columns where possible. This was done with the help of Google's geocoding API as it provides the most reliable results in terms of location. Once this data was cleaned, it was then put in tableau software to provide visualizations. In terms of country following is the mapped graph that can provide a clear picture of the location of the retweeters.

---

[1] Translation: On the Moon
[2] Translation: In Space

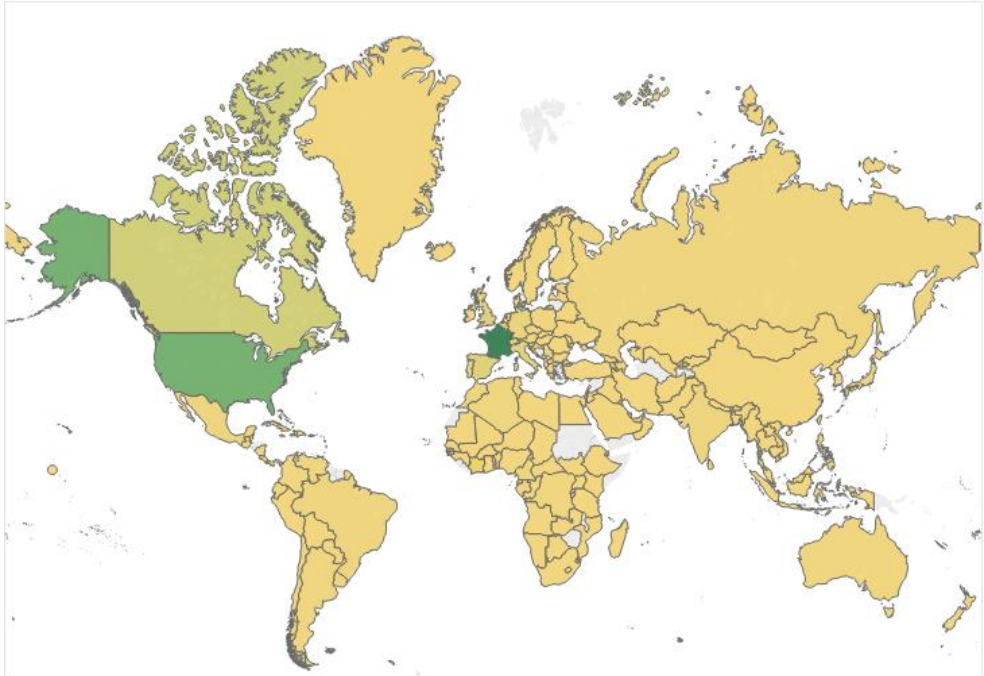

Diagram 1.6: Countries where retweeters are located

| Country | Percentage of Retweeters |
|---|---|
| France | 76.71% |
| United States | 7.87% |
| Canada | 1.98% |
| Spain | 1.26% |
| Belgium | 1.22% |
| United Kingdom | 1.15% |
| Italy | 0.90% |
| Switzerland | 0.71% |
| Germany | 0.56% |
| Morocco | 0.28% |

Table 1.4: Top 10 locations of retweeters. (Percentages have been calculated from the total of the population who reported their location)

.

More than 75 percent of retweeters reported their location to be France, which is very consistent with the language data analysis. After that, the united state is the second most popular place among the retweeters. Surprisingly the contribution of European countries toward the retweeters is lower than the US and Canada. Just to confirm that it is not due to the smaller size of European countries as compared to states, I grouped European countries to create a block (excluding France). In that case, the result came out to be that the Percentage of retweeters living in European countries outside of France is 7.3 percent of total retweeters who have given their location, as opposed to 7.87 percent retweeters from the USA. This means that there is a definite interest in retweeting French politicians from the United States. Not necessarily by American Population but this could be due to large French diasporas in these countries.

Looking into the distribution of multiple politicians among countries can perhaps provide a clearer picture of where are the foreign retweeters located. Here are the retweeters of each of politicians according to their location.

| Country | Fillon | Hamon | LePen | Macron | Melenchon |
|---|---|---|---|---|---|
| France | 82.18% | 81.23% | 69.20% | 76.61% | 74.53% |
| United States | 6.60% | 6.97% | 11.19% | 6.82% | 8.48% |
| Canada | 1.92% | 1.95% | 2.67% | 1.77% | 2.84%s |
| Belgium | 0.97% | 1.09% | 1.01% | 1.30% | 1.78% |
| United Kingdom | 0.70% | 0.86% | 1.72% | 1.27% | 0.94% |
| Switzerland | 0.64% | 0.57% | 1.04% | 0.62% | 0.86% |
| Spain | 0.54% | 0.68% | 1.20% | 0.84% | 2.30% |
| Italy | 0.38% | 0.31% | 2.44% | 0.52% | 0.42% |
| Côte d'Ivoire | 0.32% | 0.12% | 0.29% | 0.40% | 0.17% |
| Lebanon | 0.27% | 0.09% | 0.31% | 0.18% | 0.09% |
| The Democratic Republic of the Congo | 0.26% | 0.15% | 0.26% | 0.35% | 0.13% |
| Germany | 0.26% | 0.40% | 0.86% | 0.71% | 0.44% |
| Réunion | 0.22% | 0.17% | 0.17% | 0.22% | 0.25% |
| Morocco | 0.20% | 0.27% | 0.18% | 0.32% | 0.56% |
| Cameroon | 0.18% | 0.07% | 0.18% | 0.20% | 0.13% |

Table 1.5: Percentage of retweeters with respect to countries using the 'location' tab in the Twitter profile

As it can be seen from the table 1.5 that Marine Le Pen has the highest rate of retweets from the United States and she also has a very significant retweet rate from Italy (2.44 percent) as compared to other politicians, which confirms the findings from the language section of the retweeter analysis. However, here we can see that the percentages of Russian retweeters for Marine Le Pen are although the highest compared to other politicians (0.81 %), yet not very significant in numbers. It

**Department Analysis of France**

Looking into the county level data in France, we will treat the county (*Département)* as the unit of analysis. In order to have a cursory look at the data, I will first put map out the retweeters

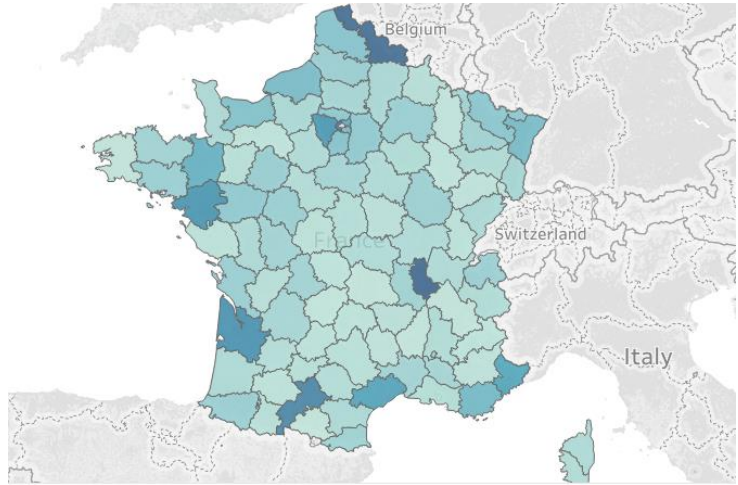

Diagram 1.7: Number of retweets located in each of the counties in France (Darker blue indicates more retweeters)

Diagram 1.7 shows the number of retweeters located in each of the counties (*Département)* in France. After looking at the map, it is clear that the retweeters are concentrated in counties with bigger cities like Marseilles, Lyon, Toulouse, Lille, Montpellier and other major French cities with large population concentrations. Although it is not very surprising, it will still be interesting to plot each of the (departement) county's number of retweeters against its population to try to see if there is a strong linear correlation.

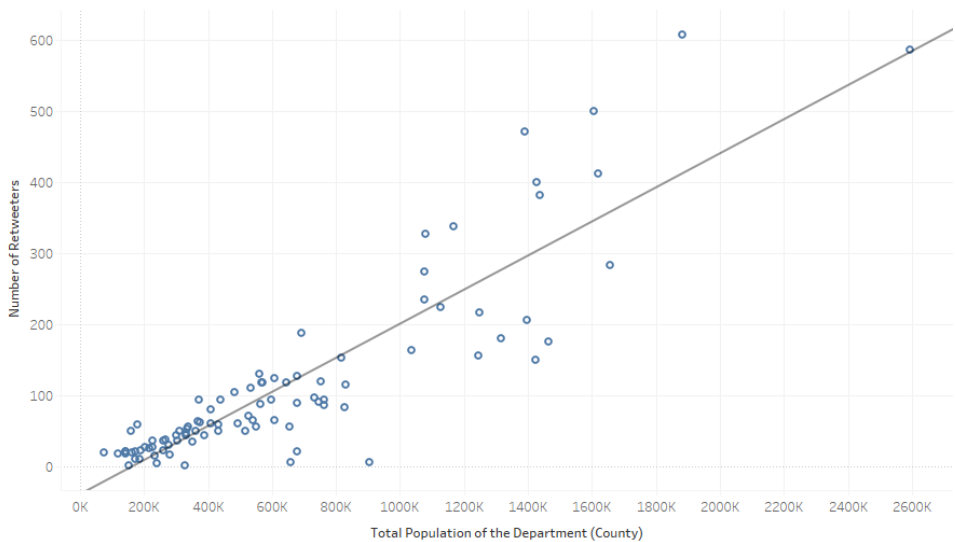

As expected, there is indeed a strong correlation between the population of each of the county and the number of retweeters it produces. The R-Square value of this regression is 0.79 which indicates a very strong correlation. There is however an exception of Paris city, which holds disproportionally more retweeters than its population, therefore Paris city was excluded from the above two graphs and needs to be treated separately. From this analysis, it seems like most of the retweeters come from large cities in France, it will, therefore, make sense to find out the percentage of retweeters that comes from the top 10 major cities in France with respect to population.

Grouping the top 10 cities of France together, it seems that 41 percent of retweeters come from these metropoles with a population larger than 200000 people. Combining this with the regression results that we have above, it will be reasonable to think that a large proportion of French retweeters live in large cities. The same experiment was replicated for each o\f the politicians separately and it was found that the results were very similar to the combined results presented above.

**Education in the Areas where retweeters are located.**

Linear correlation between population and retweeter count has some implications that we need to be careful about while analyzing the data in the next step. As the size of the population can affect the number of retweeters in the county, therefore when we are looking at the kind of areas where these retweeters come from it will make sense to take the percentage of the demographic measure rather than the actual number from the demographic surveys. For example, since the population has a high correlation with both numbers of retweeters and other measures like the number of people with higher education. In order to reduce the bias in the data, we should take the percentage of highly educated people from the total population and find its correlation with the retweeted count in that area.

An important question that one can ask about the areas where the retweeters come from is the education level and try to see if there is any relation between the education level in a department and the number of retweeters that come from that area. Data for education level in each county was collected from Insee (Insee, 2015). In each department, we looked at the level of education in terms of the percentage of the total population and then saw how it relates to the number of retweeters.

First, I compared the level of the number of retweeters in each departement with the percentage of people with no diploma or a technical studies diploma. Here are the results for this:

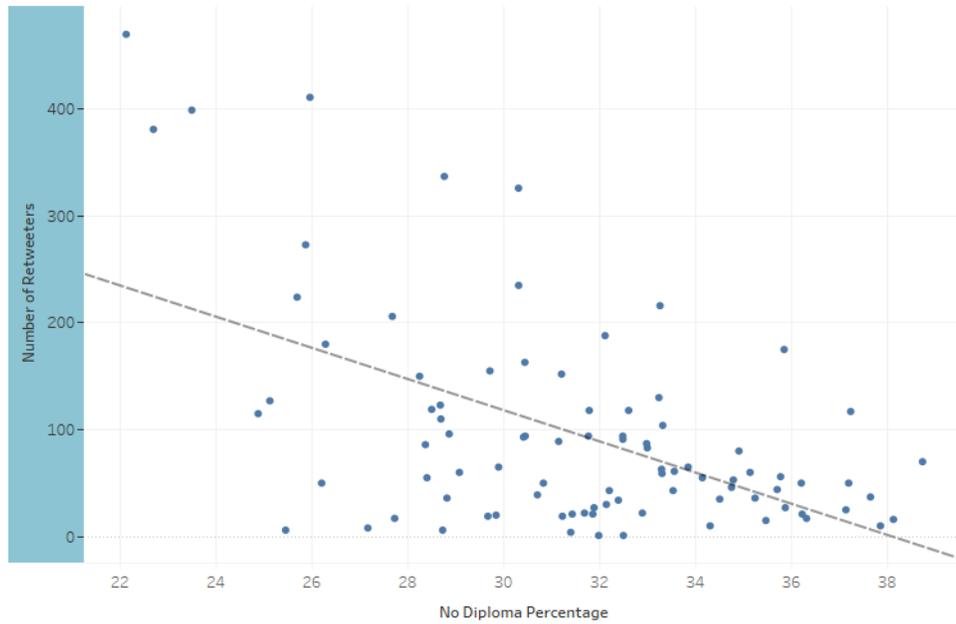

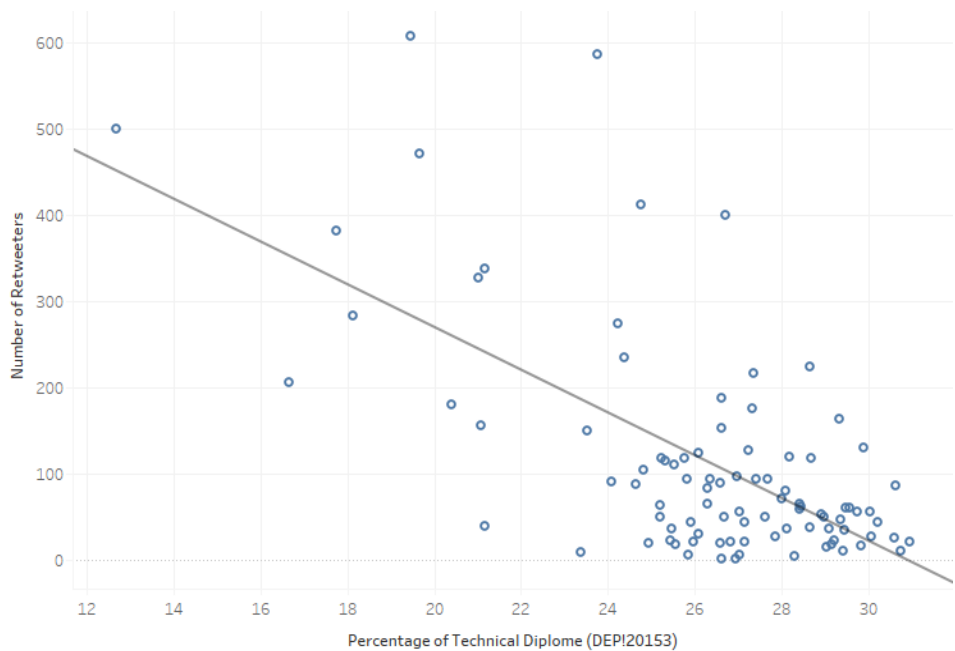

Diagram 1.8 – 1.9: These two graphs represent the relation between the number of retweeters in a departments with the percentage of no diploma holders or technical diploma holders

As visible from the above two graphs the relationship between the two variable seems to be negative. R-squared value for the no-diploma graph is 0.29 whereas it is 0.41 for the second graph with both p-values being less than 0.0001 which indicated that there is a good chance that there is a negative relationship between these variables.

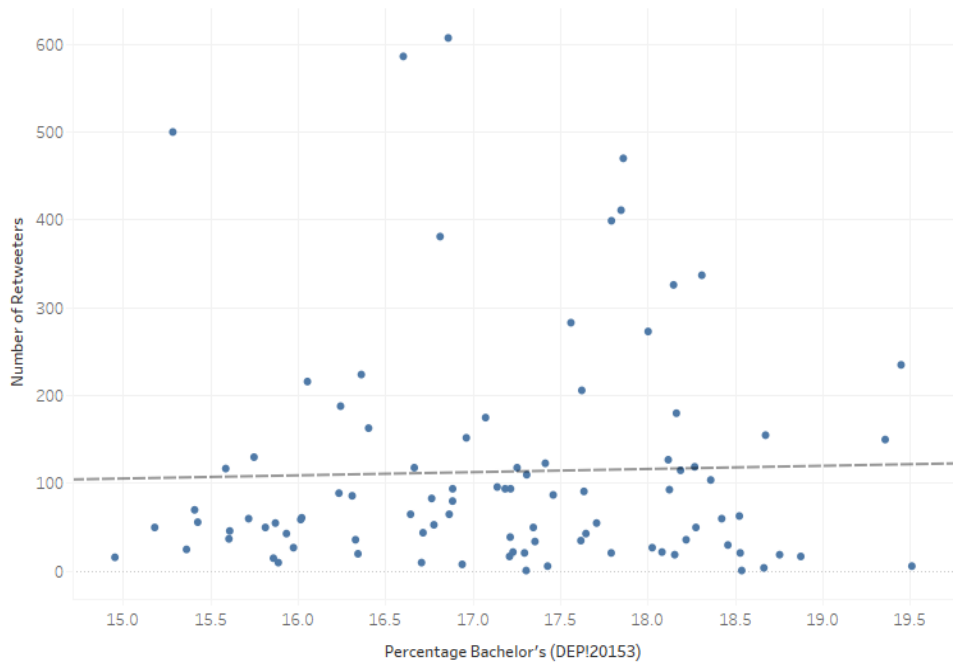

Diagram 2.0: Relation between the percentage of bachelor's degree holders and retweet count of each department

Diagram 2.0 represents the graph between the percentage of bachelor's degree holders and its relationship with the number of retweeters living in a department. The R-squared value of this graph is 0.0009 and the p-value is about 0.77 which is enough to confirm that no statistically significant correlation between the two variables is found. Although, it is observed here that the trend is moving towards positive as compared to trend from the no-diploma regression.

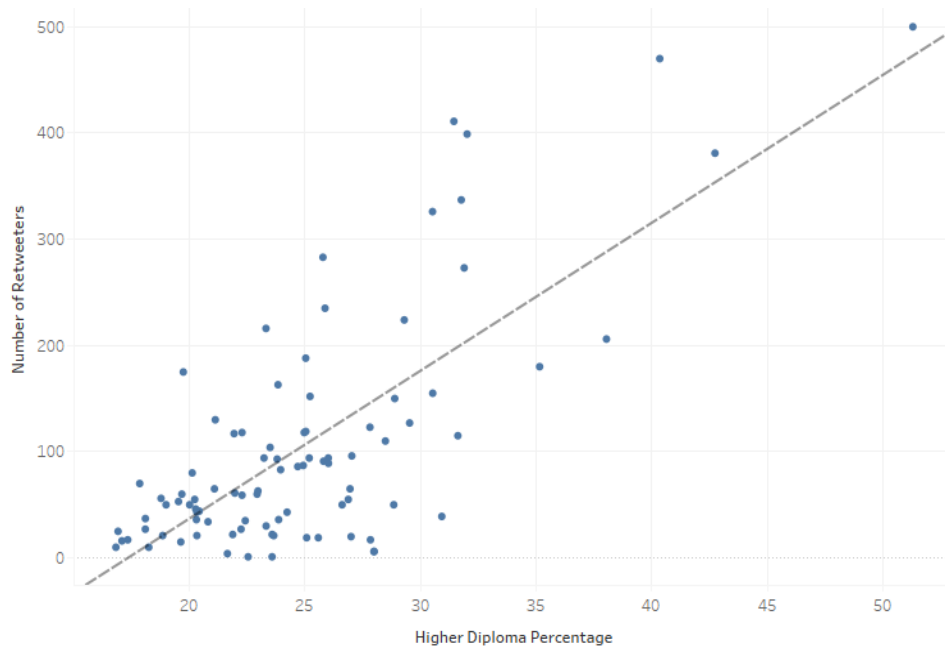

Diagram 2.1: Here I show the relationship between retweeters and the percentage of masters or higher-level diploma holders in each department.

In Diagram 2.1 we can see that there seems to be a positive correlation between the percentage of graduate or post-graduate degree holders and the number of retweeters. Here the R-squared value is 0.51, which indicates that the possibility of correlation is very high as 51 percent of data-points can be explained with the correlation. P-value is less 0.0001 here which means that there is a very little chance that this happened by coincidence.

As mentioned above, before drawing any conclusions about political retweeting and its connection with population and education, we need to study these trends in non-political groups to see if this is a general trend among all Twitter users or specific to political retweeters.

**Education in Group 2 (Political-non Retweeters)**

In order to find out the relationship between education and twitter accounts that are interested in politics but not active retweeters, the details on location were converted to the department level data using Google geocoder API from group 2. This data was then mapped in contrast to the education data from the insee with respect to each department.

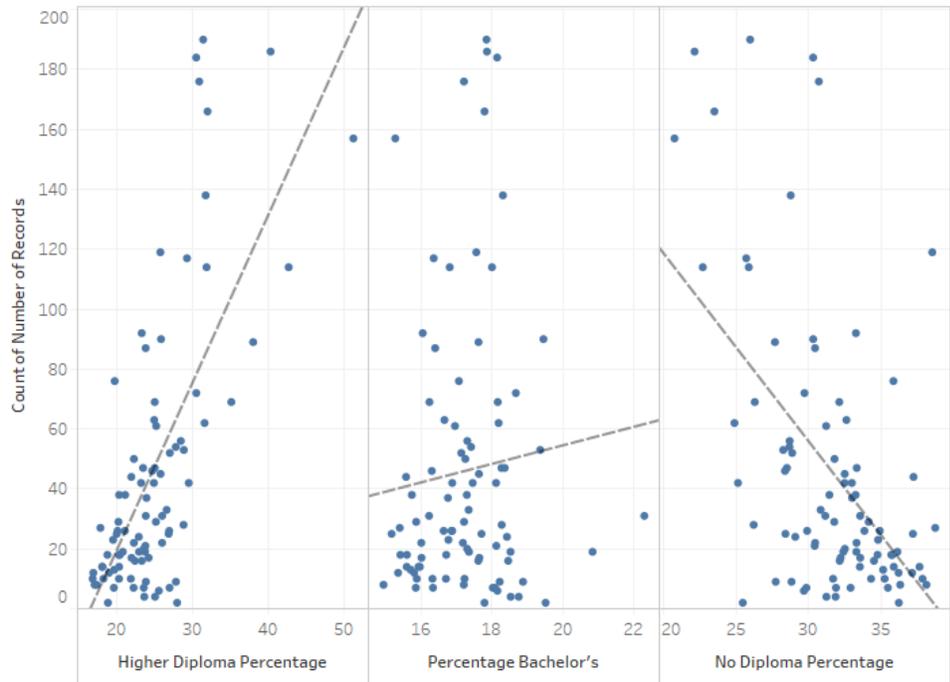

|  | R-Squared | P-Value |
|---|---|---|
| Percentage Bachelor's | 0.007 | 0.4 |
| Higher Diploma Percentage | 0.48 | <0.0001 |
| No Diploma Percentage | 0.28 | <0.0001 |

It can be clearly seen here that the trends are very similar to those of political retweeters. Although the R-values are lesser but still very close to the original number. The evidence points to the similarity between the areas where political retweeters originate from and where followers of politicians come from in terms of education.

**Education in Group 3 (Non-Political Retweeters)**

As mentioned above there are two subgroups in non-political retweeters (Footballers and Comedians). We will look at each of these groups and try to determine if the retweeters of this group are also coming from departments where the education level is higher. If that is the case then we can conclude that there is a high chance that this is the case of Twitter users in general and has no statistically significant relation with political retweeting.

Following are the graphs for the entertainment industry retweeters with respect to higher education percentage in a Département in contrast to the number of retweets coming from that Département.

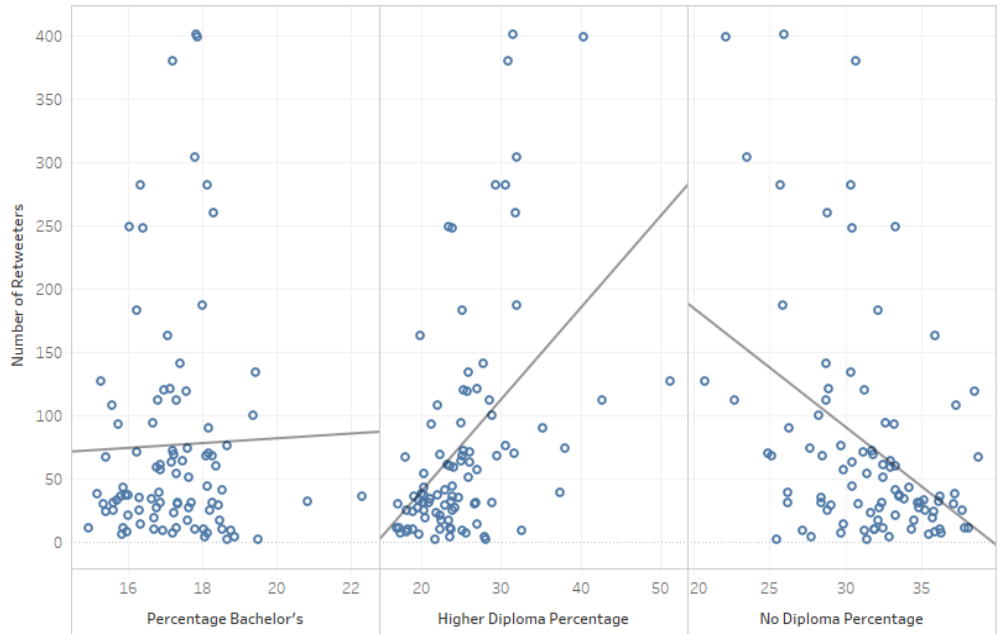

|  | R-Squared | P-Value |
|---|---|---|
| Percentage Bachelor's | 0.00069 | 0.8 |
| Higher Diploma Percentage | 0.22 | <0.0001 |
| No Diploma Percentage | 0.17 | <0.0001 |

From this graph above it is visible that the general trend looks very similar to the political retweeters. As we can see that in the departments with a higher level of education contain more retweeters from the entertainment industry then the departments with a lower level of education. But if we look at the R-squared values for each of these graphs. It is very clear that these values are much lower than group 1. For example, the R-squared values for group 1 in Higher Diploma Percentage is above 0.5, whereas it is 0.22 here. This difference is enough to consider the possibility although entertainment group retweeters come from the more educated areas in France political retweeters have a comparably higher correlation with a high-level diploma than entertainment group retweeters.

Checking the retweeters of footballers with respect to education yield the following results.

|  | R-Squared | P-Value |
|---|---|---|
| Percentage Bachelor's | 0.001 | 0.7 |
| Higher Diploma Percentage | 0.33 | <0.0001 |
| No Diploma Percentage | 0.18 | <0.0001 |

The trend in retweeters of footballers is also very similar to the retweeters of the entertainers. All the graphs look very similar to those of graph 1 and general trends are in a similar direction. But the R-squared value from political retweeters is considerably higher than both retweeters of entertainers and soccer players. It can, therefore, be seen that political retweeters are coming from areas with higher levels of education then retweeters of most popular non-political groups like entertainment industry and sports.

As it can be seen from the above-mentioned observations, that there is a positive correlation between the percentage of educated people in a department and number of Twitter profiles that come from there for each Group 1, Group 2 and Group 3. Although it is seen that this correlation becomes stronger for both political retweeters and political non-retweeters as compared to Group 3 which constitutes of retweeters of non-political personalities. This leads me to conclude that, twitter population, in general, comes from areas which have a higher level of education, but this becomes particularly true for Twitter users (Both retweeters and non-retweeters) who are interested in politicians. About political retweeters, we can say that they come from areas that are well educated, but this is not specific to just this group and we cannot conclude that they are exceptional in terms of education when compared to political non-retweeters.

**What can we know from names?**

An important and very useful piece of data that we can gather from the profiles of retweeters is the name. Names can be used as a reasonable indication of gender and ethnicity of retweeters.

How was the name data collected and cleaned?

Although twitter API does provide names of the users as entered in their profiles but a similar problem as the location was encountered, where names were either meaningless or incomplete for a large portion of the population. All the given names of the retweeter profiles were collected in a single database and then these names were partitioned into first names and last names. The first names were used for the classification of gender and ethnicity and last names were used only for the classification of ethnic background. To make this classification into subcategories based on gender and ethnic background, I needed a reference database to make the comparisons with the retweeter database. This reference database was separately created using the most frequently used baby names in the world from multiple online sources that listed baby names. All together a database of 75 thousand names database was created where each name was assigned a gender and ethnicity in which that name is most popular. This name table was then cross-matched with the retweeters database first-names to determine the gender of the profile and then with last names to determine the ethnic background of a profile.  Out of a retweeted database of 44,536 profiles, I was able to classify 23775 based on ethnic background and about 20855 retweeter profiles based on their gender using their first names.

**Gender Results for Group 1**

  From the retweeter first names. The most important information I was able to retrieve was the gender of the users. As mentioned above, it was not possible to determine the gender for all the users as many of the names were not real and it was not possible to classify them. Here is the basic gender division of the total population of retweeters.

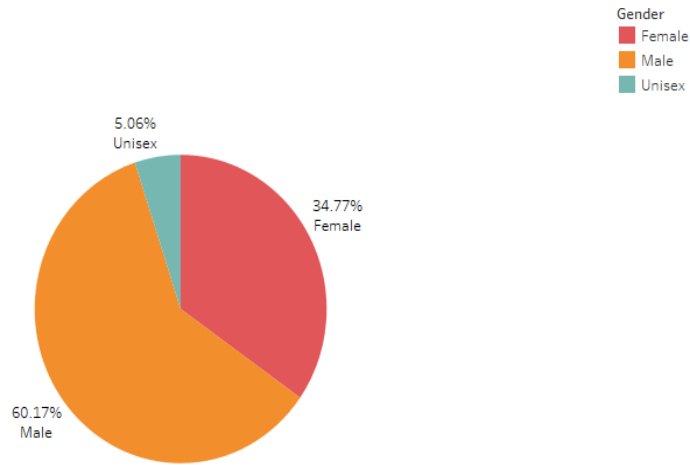

Diagram 2.3 : Gender division of the total population of retweeters.

One thing that needs explanation in this graph is the 'Unisex' label. These labels were given to the names that are usually given to both boys and girls. Therefore, it was difficult to classify them. As seen from the graph, a large portion of retweeters is male (60 percent) and the female population represents only 34 percent of the retweeters. It will, however, be interesting to see if this male-female ration is maintained when we separate the French retweeters from the foreign retweeters. This separation will be done using the country tag which we have found out in the location section of this chapter. Here is the result for retweeters from France.

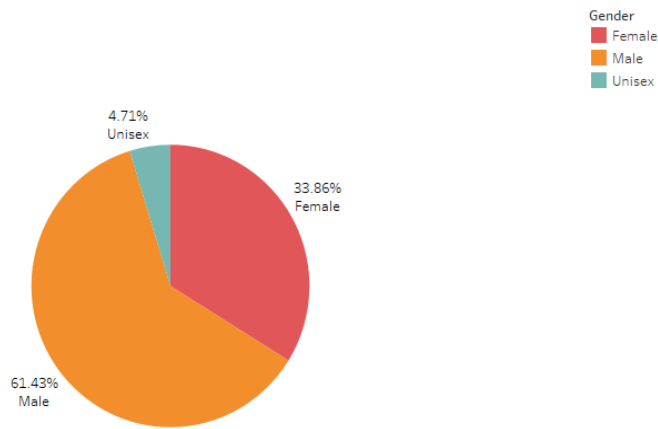

Diagram 2.4: Division of gender of retweeters in France

As seen from the diagram 2.4 the division of gender for French retweeters is roughly the same as all retweeters. Percentage of male retweeters has only increased 1.43 percent at the expense of female retweeters and unisex names. However, the general trend is the same as above. It is, therefore, possible to infer that the French Political retweeting is generally male-dominated. The analysis of gender will, however, be incomplete without looking into the candidate individually to see if there is a different picture on that level.

**Gender Bias on individual Candidate Level**

Here is a table describing the gender distribution of each of the individual politicians in France.

|            | HAMON  | MELENCHON | LE PEN | MACRON | FILLON |
|------------|--------|-----------|--------|--------|--------|
| **FEMALE** | 35.87% | 35.36%    | 35.02% | 33.66% | 31.52% |
| **MALE**   | 59.22% | 57.66%    | 60.64% | 61.98% | 63.61% |
| **UNISEX** | 4.92%  | 6.98%     | 4.35%  | 4.37%  | 4.87%  |

Table 1.7: Individual gender distributions for each of the politicians

In general, it can be observed that the gender distribution for each of the politicians is very close to the collective gender distributions. Although, some deviations can be observed here. For example, Francois Fillon has considerably larger male retweeters when compared to the total male retweeters percentage (60 Percent), He also has a smaller percentage of female retweeters compared to other politicians and collective gender distribution from table 1.7.

Another anomaly that is boggling me, is that Melancon has a much lower percentage of male retweeters than others which becomes puzzling as a percentage of his female retweeters is the same and yet the percentage of unisex retweeter names are much higher for him. It was therefore manually checked to see if this was through an error in processing and it was discovered that this reflects the data correctly and no processing error has been made at this stage.

**Ethnic Background and political retweeting**

Because of prohibition in France to collect ethnicity related data, it is very difficult to find data on this matter. Here the context of this data processing is that we discovered in the language exploration of the retweeter data above that Arabic represented a small portion of retweeters as opposite to a significant presence in French population according to data from OECD. It is, therefore, necessary to countercheck if this is indeed the case or there is some error that has been made in the data processing. This was done by finding out the origin of the Twitter names. It is reasonable to assume that name can be used as a good measure to make an assessment of the presence of ethnic diversity among the retweeters. Simons has shown that in the absence of ethnic data, the next best thing to use is name (Simon, 2010).

**Results for the ethnic inquiry**

It is found that the Arabic names represent only 3.05 Percent of the total population and this result is very far from the result we got for the Arabic language 0.05 percent and much closer to OECD data which showed that people who speak Arabic in France represent roughly 3 percent of the population. It was found that most of the people with names of Arabic origin live in France and tweet in French. This confirms my first explanation from the language inquiry that Arabic speaking people in France are more likely to tweet in French instead of using Arabic as their interface language for twitter. So, we can conclude that the idea of underrepresentation of Arabic people in political retweeting is not based on facts.

**Gender Results for Group 2**

The same process as above was repeated for the non-retweeters who are interested in politics. This was done to put the above findings in perspective. Following results were found from this inquiry.

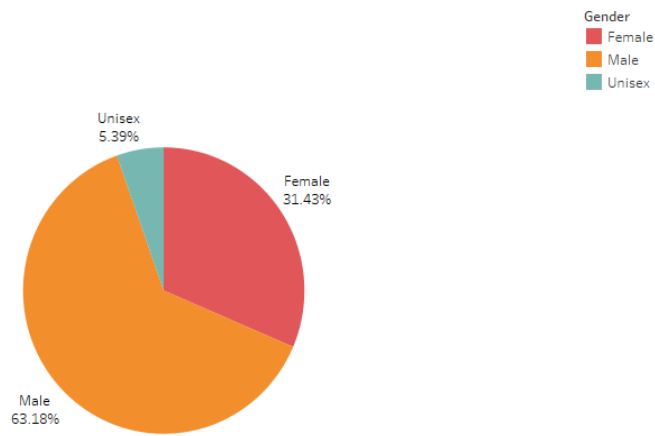

Diagram 2.4 Gender breakdown of Group 2

**Gender Results for Group 3**

The same process as above was repeated for the non-political retweeters. Following results were found from the inquiry of entertainer retweeter group.

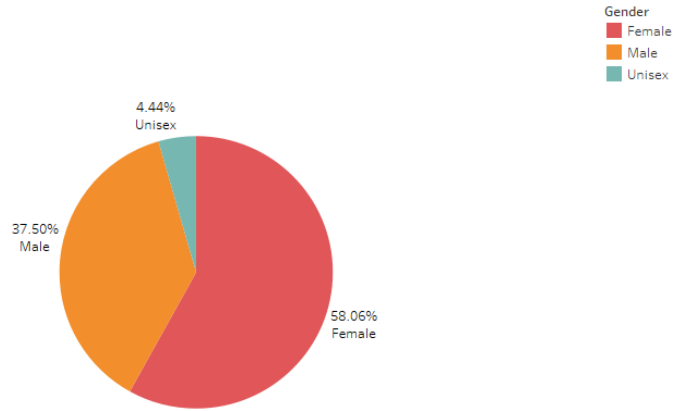

Diagram 2.5 : Gender breakdown of entertainment personality retweeters

The footballer retweeters yielded the following result with respect to gender.

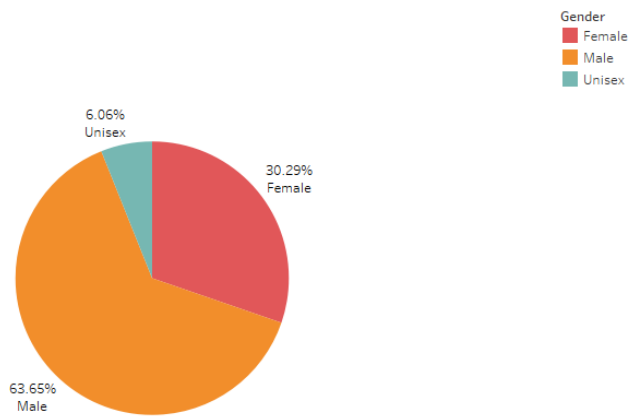

Diagram 2.6 : Gender breakdown of football retweeters

**Conclusions about political retweeters from gender-related category**

It can be seen from the graphs above that the gender division is very uniform between (Group 1) political retweeters and (Group 2) political non-retweeters. However, when we look at the non-political retweeters we can see from the difference between the gender divisions of retweeters of footballers and retweeters of entertainers that gender division is something very particular to each group and not a general trend among all groups on twitter.

**Self-Description of Retweeters**

This analysis will not be complete without coming up with a generic picture representing the self-description and professions of political retweeters. It is an important factor in this analysis as a cursory look in the data reveals that many of the retweeters have a professional interest in retweeting the politicians. They are either politicians themselves or then involved in political journalism of some sort, it will, therefore, be interesting to verify this from the data. For the purpose of this inquiry, I will divide the data into 8 broad categories.

Politicians

Official Political Party Accounts

Political Party title holders

Government Officials or Government Offices

Political Journalists and Bloggers

Academic personals, (Students, Researcher, Teachers)

Generic Description profiles

Unidentifiable

**Categorization of Retweeters according to the self-description**

Professional identification using Twitter is an extremely difficult task as there is no profession related section in twitter API that will reveal that information about an account. Two approaches were explored to detect this kind of information. The first approach included the cross-matching of twitter profile with LinkedIn profiles as LinkedIn is a more professional platform where one is more likely to find professional information. This approach turned out to be unsuccessful as only 10 percent of retweeter profiles could be found on LinkedIn and this ratio was not good enough to detect the general trend. The second approach that was tried was the usage of twitter description to identify the professions, which turned out to be relatively more successful. In our case, 72 percent of the retweeters have written something in their description while others have left that section blank. For each of the above-mentioned categories, certain keywords were selected that can be used to describe that category of individuals. A generic search for these keywords in the self-description database revealed the professional inclinations of retweeters. Following is the list of keywords searched for each of the categories.

| Profession | Keywords Searched | Verified | Profiles Found | Percentage |
|---|---|---|---|---|
| Politicians | Ministre, Député, Dpute, Politician, senateur, Snateur, | no | 878 | 1.9% |
| | Maire, Prsident, Commissaire, | | | |
| Political Party title holders | | | 1455 | 3.25% |
| Government Offices | Compte officiel , Ambassade, | Yes | 68 | 0.15% |
| Political Journalists and Bloggers | Blog, Journal, Tele , auteur, écrivain, rédacteur , commentateur , critique, éditorialiste | No | 1158 | 2.5% |
| Academic personals | Etudiant, prof, professeur, chercheur, ecole, linguist, science po, universit , scien, phd , docto , Diplôme , License, lyce, student | No | 2085 | 4.66% |
| Political Party and group Accounts | Fillon, officiel, macron, republicains, Jeunes Républicains, en marche, em , ump, jeunes avec, parti socialiste, socialiste, hamon, ps, Fn, Front National, marine, pen, mlp, melechon, insoumis | No | 523 | 1.1% |
| Dedicated Supporter Accounts | | No | 4684 | 10.47% |
| Generic self-description with no explicit political adherence | | No | 24021 | 53.7 % |
| No Description | | No | 11915 | 25.03% |

Since the data about education and salary is not directly reported by the retweeters and is based on secondary information collected from their profile, we must be very careful in drawing conclusions from this information.

Correlation cannot certainly be considered the same thing as causation and even if there is a healthy correlation between the retweeters count, what can conclude from this correlation is that people living in departments with higher education level are more likely to retweet French politicians than the people who live in areas with lesser number of highly educated people.

**Conclusions**

Exploring the demographics of political retweeters has made it clear that demographics of political retweeters are generally very close to demographic results found by Barbera is largely true for political retweeters more than the non-political retweeters (Barbera & Rivero, 2014). The space of political retweeting is male dominated and concentrated in urban areas with relatively higher education levels. It has also been revealed that a very significant percentage of political retweeting happens on accounts of dedicated political workers and party accounts, which corroborates with the literature whereas the presence of international population among both political and non-political retweeters is highly representative of real-world international popularity.